\documentclass[sigconf,screen,author]{acmart} 

\AtBeginDocument{%
  }

\usepackage{mwe}

\copyrightyear{2026}
\acmYear{2026}
\setcopyright{cc}
\setcctype{by}
\acmConference[ASE '26]{Proceedings of the 41st IEEE/ACM International Conference on Automated Software Engineering}{October 12--16, 2026}{Munich, Germany}
\acmBooktitle{Proceedings of the 41st IEEE/ACM International Conference on Automated Software Engineering (ASE '26), October 12--16, 2026, Munich, Germany}
\acmDOI{10.1145/3832783.3834651}
\acmISBN{979-8-4007-2882-2/2026/10}

\begin{document}

\title[Cochise: A Reference Harness for Autonomous Penetration Testing]{Cochise: A Reference Harness for\\ Autonomous Penetration Testing}

\author{Andreas Happe}
\email{andreas.happe@tuwien.ac.at}
\orcid{0009-0000-2484-0109}
\affiliation{%
  \institution{TU Wien}
  \city{Vienna}
  \country{Austria}
}

\author{Jürgen Cito}
\email{juergen.cito@tuwien.ac.at}
\orcid{0000-0001-8619-1271}
\affiliation{%
  \institution{TU Wien}
  \city{Vienna}
  \country{Austria}
}


\begin{abstract}

Recent work on LLM-driven autonomous penetration testing reports promising results, but existing systems often bundle architectural, prompting, and tool-integration choices together. This makes it difficult to determine what is gained over a simple agent and harness. We present Cochise, a 630 LOC Python reference implementation for autonomous penetration-testing experiments. Cochise connects to a Linux execution host over SSH and supports attacking controlled target environments reachable from that jump host.
The prototype implements a Planner--Executor architecture in which long-term state is maintained by the planner, while a ReAct-style executor issues commands over SSH and self-corrects based on command outputs. The scenario prompt can be adapted to different target environments. We evaluate the harness against a live third-party testbed, Game of Active Directory (GOAD).

Cochise is intended not as a state-of-the-art penetration-testing agent, but as a reusable experimental infrastructure for comparing models, agent architectures, and penetration-testing traces.

Alongside the prototype, we release replay and analysis tools: (i) cochise-replay for offline visualization of captured runs, (ii) cochise-analyze-logs and cochise-analyze-graphs for cost, token, duration, and compromise analysis, and (iii) a corpus of JSON trajectory logs from GOAD runs, so that researchers can study agent behavior without provisioning the 48--64 GB RAM / 190 GB storage testbed themselves. Tool demo video available at \url{https://youtu.be/2mQimB1ufyI}.
\end{abstract}




\received{20 February 2007}
\received[revised]{12 March 2009}
\received[accepted]{5 June 2009}

\maketitle

\section{Introduction}

LLMs for autonomous network penetration testing have drawn both academic and industrial interest. Publications report performance numbers for their respective prototypes, but reusable execution and logging infrastructure is typically missing. Comparing models and architectures under a common protocol or reusing traces and analysis pipelines across studies is therefore difficult. Cochise is a minimal reference agent and harness\footnote{A small, stable scaffold/harness against which alternative designs can be ablated.} rather than a high-performance penetration-testing agent. It provides an execution interface, model abstraction, state handling, and unified trajectory format so that researchers can compare models, architectural variants, and agent traces under a common protocol. By using the LiteLLM framework, it allows for easy exchange of the underlying LLM. Cochise's configurable scenario prompt allows integration into diverse use-cases. At 630 lines of Python, the codebase is small enough to read end-to-end, and thus can serve as a starting point for new prototypes or as a baseline for benchmarking.

\paragraph{Provided Artifacts}

We provide the \verb|cochise| autonomous penetration testing tool, and \verb|cochise-replay| for offline visual replay of captured log files. For log analysis, \verb|cochise-analyze-logs| summarizes collections of trajectory logs by model, token utilization, cost, and success rate, and exports the results as LaTeX-tables. \verb|cochise-analyze-graphs| allows generating graphs.

We also release a collection of penetration-testing trajectory logs to enable post-hoc analysis without incurring token costs.

\paragraph{Intended Users}

The primary audience are researchers who need a controllable and minimal prototype for studying autonomous penetration-testing agents. They can reuse Cochise's SSH execution interface, trajectory format, replay tooling, and analysis scripts; compare LLMs through the LiteLLM backend; or analyze the released traces without running the live testbed.

\section{Motivation, Background, and Related Work}

There has been high academic interest in using LLMs for autonomous penetration testing, with current prototypes including incalmo~\cite{singer2025incalmoautonomousllmassistedred}, cAI~\cite{mayoralvilches2025caiopenbugbountyready}, and pentestGPT\footnote{We include pentestGPT although it is single-host and not multi-host centric.}~\cite{deng2026makesgoodllmagent}. They commonly focus on raising their respective success rates, provide multiple operation modes, and are under continuous development. PentestGPT used a prior version of our work~\cite{10.1145/3766895} as a point of comparison, which we read as one indication of demand for reusable baseline prototypes combining a minimal agent and scaffold. We therefore wrote \texttt{cochise}, a minimal baseline harness/agent with explicit execution, logging, replay, and analysis interfaces.

\paragraph{Penetration Testing as an SE Problem}

While our target domain is security-centric, the problem an autonomous penetration-testing agent must solve originates in the software engineering domain.
Realistic security evaluation testbeds are dynamic stateful environments where actions can trigger destructive side-effects: invoking an exploit can crash the target system, enumeration actions can trigger intrusion detection systems, and password-based attacks can lock out accounts. Automated EDR solutions act as active defensive adversaries, making the scenario an attacker/defender setup.

Agents must cope with these side-effects and combine findings into successful attack paths, making their trajectories inherently multi-step. The agent can run arbitrary commands on the execution host (Section~\ref{system_overview}), which leaves the action space unbounded. Penetration-testing tools are also fickle and often complex to use, so the agent needs robust error handling and auto-repair.

Partial observability, side-effecting actions, an unbounded action space, and the need for autonomous error recovery also characterize a broader class of agentic SE tasks~\cite{yang2024sweagent}. We therefore study autonomous penetration testing as an instance of agentic software engineering rather than as security tooling alone.

\paragraph{Related Harnesses}

Architecturally, Cochise sits between minimalist agent scaffolds such as mini-SWE-agent~\cite{yang2024sweagent} and full-feature penetration-testing agents such as cAI~\cite{mayoralvilches2025caiopenbugbountyready} and incalmo~\cite{singer2025incalmoautonomousllmassistedred}. The closest minimalist comparison, mini-SWE-agent, lacks long-horizon trajectory management. In pilot runs against our target, GOAD, it could perform initial exploitation but did not progress to multi-stage compromise.
We report this as an observation about one scaffold--model pair at
one point in time rather than as a property of minimal scaffolds in general.
We expect the scaffolding required for a given task to shrink as model capability increases, and discuss how this affects the benefits of bespoke scaffolds in Section~\ref{sec:discussion}.

\section{Architecture}

Cochise operates in an assumed-breach setting. Emulating an attacker that has gained access to an internal enterprise network, it starts with command execution on a single Linux host situated within the target network. In our evaluation, this host is a Kali Linux VM\footnote{Kali Linux is a specialized Linux distribution which comes with common penetration testing tools pre-installed.} connected to GOAD, and the agent's task is to enumerate and compromise Active Directory domains. The harness itself only provides command execution over SSH and structured observation of command outputs. Cochise does not address initial access (phishing, exploit-driven remote code execution, physical access), and we do not model human-in-the-loop adversaries. The execution host is trusted; the testbed network is the the only attack surface.

\begin{figure}[h]
  \centering
  \includegraphics[width=0.8\linewidth]{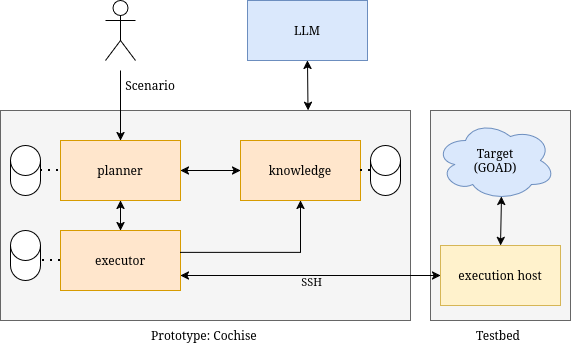}
  \caption{System and Agent Architecture. Orange components designate the Cochise prototype, blue components are third-party tools outside of Cochise's direct control. The execution host (in yellow) is a Kali Linux VM through which Cochise interacts with the target environment.}
  \Description{Diagram showing the system and agent architecture.}
  \label{sys_architecture}
\end{figure}

\begin{figure*}[h]
  \centering
  \includegraphics[width=\linewidth]{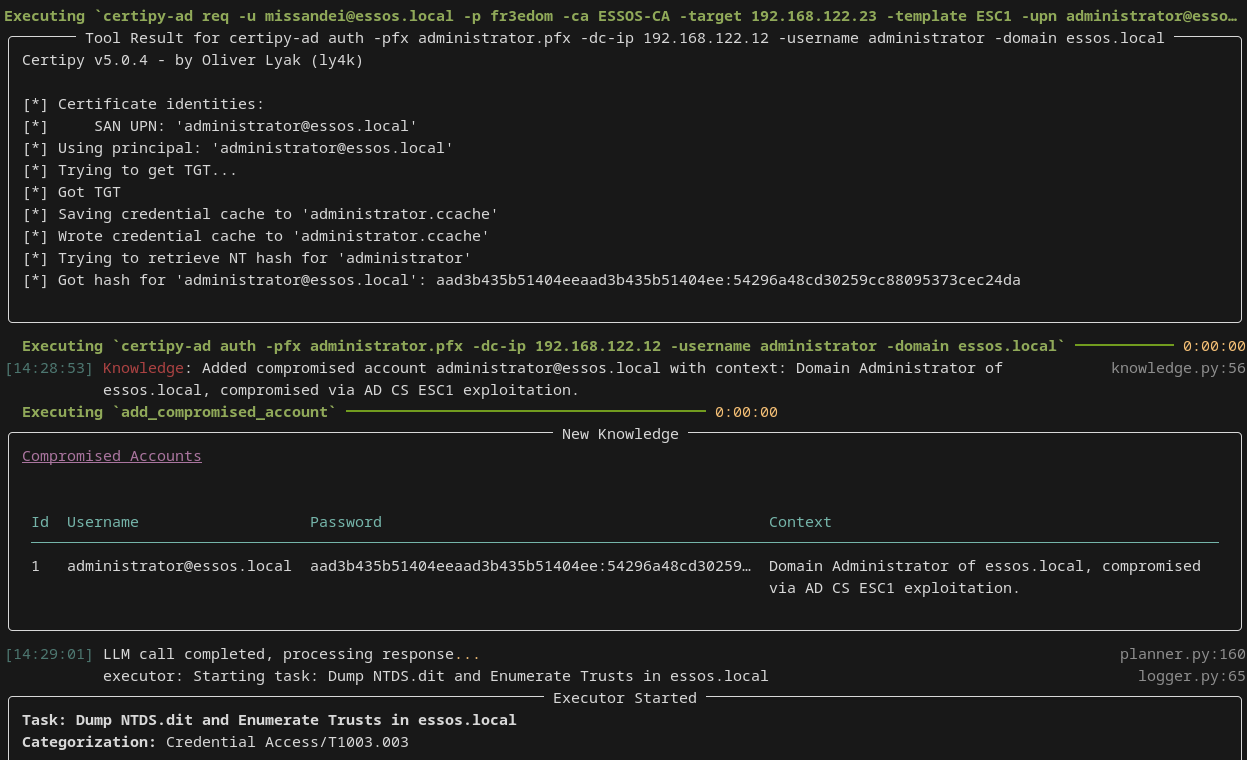}
  \caption{Screenshot of an active Cochise run, showing how Cochise uses \texttt{certipy-ad} to execute an ADCS ESC1 attack, gains Administrator level access to a domain controller, and then dumps all user credentials stored on this server.}
  \Description{Screenshot of an active Cochise run within a Terminal window, showing how Cochise uses \texttt{certipy-ad} to execute an ADCS ESC1 attack, gains Administrator level access to a domain controller, and then dumps all user credentials stored on this server.}
\end{figure*}

\subsection{System Overview}
\label{system_overview}

The overall architecture is shown in Figure~\ref{sys_architecture}. The Cochise prototype connects a cloud- or locally-provided LLM to a third-party penetration-testing testbed such as GOAD. A separate virtual machine is added to the testbed as an execution host. Cochise connects to this execution host over the standard SSH protocol and uses it to execute commands within the target network. These commands typically perform reconnaissance or attacks and are thus limited to the target network.

We deliberately keep the Cochise source code on a separate host from the testbed and connect to the testbed only via SSH. This protects against the accidental destruction of the agent's own configuration or logs by the agent itself, and prevents lateral movement back from a compromised target host to the agent. The agent can execute arbitrary commands on the execution host but cannot modify its own source or logging configuration. This also allows the testbed to be operated offline while Cochise connects online to an LLM-provider.

We use a Kali Linux virtual machine optimized for penetration-testing, but the prototype is not tied to it. A standard Linux virtual machine such as Debian, combined with a new scenario prompt, lets the prototype perform tasks outside the security domain, e.g., system administration.

\subsection{Prototype Architecture}

The architecture consists of a high-level \verb|planner| component that maintains long-term state, and a low-level \verb|executor| component that keeps only ephemeral per-task state. Both planner and executor can submit findings as well as compromised accounts to the \verb|knowledge|~component.

The \verb|planner| is the strategic core of the prototype. It maintains long-term state in a continuously updated structured textual task representation. In our default configuration, we use a Pentest-Task-Tree (PTT)~\cite{deng2026makesgoodllmagent} which is a tree-structured Markdown task list, but neither agent nor harness depend on this exact format. Both \verb|planner| and \verb|executor| operate in rounds in which the \verb|planner| designates work-packages with concrete goals and forwards them to the \verb|executor| component for solving. During each strategy round, the \verb|planner| receives the current PTT, a summary of the most recently executed task, and queries the \verb|knowledge| component for its current world-view. Using this information, it selects the next task to be executed and instantiates a new \verb|executor| with this task. After the \verb|executor| has finished, a new strategy round begins. In addition to issuing tasks, the \verb|planner| can also submit new information to the \verb|knowledge| component and revise the PTT.

The \verb|executor| is a ReAct agent~\cite{yao2023reactsynergizingreasoningacting} that translates the \verb|planner|'s high-level directives into concrete operational commands. It connects over SSH to the execution host inside the test environment and issues the commands needed to solve the task given. Once it judges the task to be completed, or a maximum number of interaction rounds is reached, it returns a summary of its invocation to the \verb|planner|. Afterwards the \verb|exeuctor| and its state are discarded. Executor memory is thus ephemeral and episodic: a new executor instance is created per task and discarded after completion. As each \verb|executor| starts with a fresh context, this bounds the executor's context window size and cost which otherwise would grow across hours of operation. It does force the planner to be the integration point for cross-task knowledge. This also prevents transient \verb|executor|-level failures from propagating to later executor invocations, e.g., contains malformed local reasoning traces, stale short-term observations, or task-specific prompt drift within a single \verb|executor| round. Discarding the executor does not reset the execution host nor undo side effects in the target environment.
While the planner--executor architecture reduces context size and experiment costs, it is not strictly mandatory for performing successful attacks against the GOAD testbed as the initial exploitation performed by the non-hierarchical \verb|mini-swe-agent| has shown. We discuss non-performance benefits of this architecture in Section~\ref{sec:discussion}.

Both planner and executor| implement the Reflexion~\cite{shinn2023reflexion} pattern to detect and repair invalid command invocations.

\subsection{Logging and Analysis}


Cochise records every interaction between the \texttt{planner}, the executor, the LLM APIs, and the target network in a per-run JSON log file in a format designed for downstream data analysis and reproducibility. An LLM invocation, for instance, is logged with an event key naming the specific architectural action that produced it, e.g., \verb|planner_task_selection|. The entry contains sub-dictionaries detailing the exact prompt submitted to the model, the raw text completion received, and per-call cost metrics including input tokens, output tokens, reasoning tokens, and cached tokens. Issued commands use separate event identifiers and record the exact bash string executed on the execution host alongside the resulting standard output and standard error streams.

These logs lower the entry cost of LLM security research. GOAD, the live environment used in our evaluation, runs five concurrent Windows Server virtual machines alongside a Kali Linux attacker execution host, and needs a minimum of 32GB to 48GB of RAM and roughly 190GB of disk storage. These hardware requirements exclude academic researchers and students from performing experiments and gathering attack trajectories. The Cochise artifact therefore ships Python scripts that parse the provided JSON logs, as well as a dedicated replay tool that can be used to display and reenact an attack run without executing its commands. With them, researchers can reconstruct the evolution of the high-level plan, visualize the command execution flow, and extract cost and token metrics without provisioning the GOAD environment. The released traces capture interactions between frontier LLMs and a live penetration-testing testbed, and can be analyzed directly or reused within secondary studies.

\section{Evaluation}

We evaluate Cochise along three dimensions: compactness, capability, and analyzability. Compactness captures whether the scaffold is small enough to inspect and modify. Capability captures whether it can produce meaningful long-horizon trajectories in a live testbed. Analyzability captures whether generated runs can be replayed and converted into reusable empirical data.

\begin{table}[t]
  \caption{Complexity as measured by the Lines-of-Code. For Core LoC, we used src/cai/\{agents, internals, prompts, tools\} for cAI, incalmo/core for incalmo, pentestgpt/\{core, prompts, tools\} for pentestGPT, and the LoC for the full attacker for cochise.}
  \label{tab:loc}
  \begin{tabular}{lrr}
    \toprule
    Prototype & LoC in Repo. & Core LoC \\
    \midrule
    cAI~\cite{mayoralvilches2025caiopenbugbountyready} & 41183 & 9100 \\
    incalmo~\cite{singer2025incalmoautonomousllmassistedred} & 52603 & 4902 \\
    pentestGPT~\cite{deng2026makesgoodllmagent} & 3779 & 1822 \\
    cochise (this work) & 1328 & 630 \\
  \bottomrule
\end{tabular}
\end{table}

\paragraph{Prototype Compactness}

We use the number of lines-of-code as coarse proxy for compactness and complexity. We count the effective lines-of-code of different autonomous penetration-testing prototypes using \verb|tokei|\footnote{\url{https://github.com/XAMPPRocky/tokei}}. To allow for fair comparison, we count both the total amount of code in the respective GitHub repositories as well as the amount of code for the prototype's core functionality, typically consisting of core logic, prompts, and provided tools.

Table~\ref{tab:loc} shows that Cochise's $630$ lines-of-code are a factor of 3--15 smaller than the other prototypes' core code. The closest alternative, pentestGPT, is built on top of Anthropic's \verb|claude-code|, whose code-base we have not included in our count. This also currently prevents usage of pentestGPT with non-Anthropic models.

\paragraph{Prototype Capability}

\begin{table}[t]
  \caption{Capability Evaluation using cochise and GOAD. \textit{Account} is a fully compromised account. Summary of 5 runs per model. All values are mean $\pm$ deviation, costs are in USD.}
  \label{tab:performance}
  \begin{tabular}{lrrr}
    \toprule
    Model & Accounts/h& Cost/h & Cost/Account\\
    \midrule
    Gemini-3-Flash  & $15.77 \pm 8.17$  & $1.75 \pm 1.03$ & $0.11 \pm 0.03$ \\
    Claude-4.7-Opus & $37.75 \pm 20.13$ & $21.35 \pm 5.93$ & $0.74 \pm 0.25$ \\
  \bottomrule
\end{tabular}
\end{table}

In order to generate usable attack trajectories, the prototype must be capable of performing successful attacks against an enterprise networks. In addition to the results presented in~\cite{10.1145/3766895}, we performed a capability evaluation using GOAD. We highlight results of two current frontier LLMs, Google's \textit{Gemini-3-Flash (preview)} and Anthropic's \textit{Claude-4.7-Opus}, in Table~\ref{tab:performance}.

Across all runs, both Opus and Gemini-Flash compromised at least one of the three Active Directory domains within GOAD; in 3/5 runs, both models were able to compromise all three domains. Gemini-Flash reached this at 12x lower average cost per hour ($\$1.75$ vs. $\$21.35$) compared to Opus-4.7. We do not interpret these results as evidence that Cochise's architecture is optimal. Instead, they indicate that Cochise is capable enough to generate meaningful long-horizon trajectories on a live testbed, and that its logging exposes cost--capability trade-offs across models.

\paragraph{Analyzability} The prototype should be usable as data source or baseline for researchers. We use adaptation by other research projects as proxy for analyzability. As we are only now releasing this prototype on GitHub, we cannot yet report third-party adoption by independent researchers. An earlier version of Cochise was used as a comparison point by pentestGPT~\cite{deng2026makesgoodllmagent}, which we count as positive precedent.

\section{Discussion}
\label{sec:discussion}

\subsection{Harnesses and Agents}

As part of research into autonomous penetration testing~\cite{10.1145/3611643.3613083,happe2026llms,10.1145/3766895,anthropic2026cybertoolkits}, complex early harnesses were progressively replaced by simpler ones paired with more capable models. This raises a question: are bespoke security harnesses a transient artifact of insufficient models, or a permanent feature of the problem domain? If capability that today requires explicit orchestration is tomorrow absorbed into the model's own long-horizon reasoning, investing time and effort into harnesses yields only limited temporary benefits. We therefore discuss benefits of scaffolds that go beyond making frontier models effective at penetration testing.

\paragraph{Enabling Small Language Models (SLMs)} \verb|mini-swe-agent| has shown that a combination of a minimal harness with a frontier-model can competitively solve software-engineering problems~\cite{yang2024sweagent}. Similar experiences have been reported for frontier models in the cybersecurity domain~\cite{anthropic2026cybertoolkits}.  While this is a plausible argument for frontier models, it is not for the field as a whole. Data privacy and digital sovereignty concerns are driving the adoption of smaller, cheaper, and open-weight models. Prior work reports that explicit harness abstraction layers improve the performance of smaller models on multi-host tasks~\cite{yang2026betterharnessessmallermodels,singer2025incalmoautonomousllmassistedred}.

\paragraph{Harness providing Structure.} A harness imposes structure on otherwise free-form behavior. In Cochise, the planner selects units of work with a defined goal and context, and adds a MITRE ATT\&CK classification to aid later analysis. After the executor completes each unit, it returns a structured record of what was executed and what knowledge was gained. This imposed structure aids context management, makes runs more comparable, and supports more detailed analysis of the agent's performance. Cochise also labels every LLM invocation with the architectural action that produced it, so input, output, reasoning, and cached tokens are attributable to distinct features, e.g., strategy revision or error recovery. Questions such as ``what fraction of budget goes to re-planning?'' or ``does reasoning-token spend concentrate in recovery from failed commands?'' thus become directly answerable from the logs.

\paragraph{Harness providing Safety.} The harness mediates every interaction between agent and environment, which makes it the natural place to enforce safety constraints. At the interaction level, it can log and gate all actions requested by the agent, and the agent's capabilities are bounded by the actions the harness provides. Oversight can be added by either keeping humans in the loop or by using LLMs-as-Judges before executing operations.

The overall architecture also restricts potentially dangerous capabilities. For example, Cochise runs the agent on a separate host outside of the target test network. To interact with its targets, the agent can only issue commands which are then executed on a separate virtual machine within the target network. Keeping the agent's source, runtime and logs on a separate host means an agent cannot alter its own configuration, source code, or trajectory record. If cloud-provided LLMs are used, only the host running the agent needs a connection to the LLM-provider, and thus to the Internet, while the systems on the target network can be kept offline. This limits the damage that a rogue agent can perform.

\paragraph{Research Ergonomics.}

Beyond enabling the underlying model, scaffolds used in offensive agent research should support researchers. A good scaffold provides vendor independence, so that models from multiple vendors can be run under one interface. It provides unified logging and analysis, so that runs can be compared and data shared between researchers. And it gives researchers a stable base for their own prototypes, as well as a minimal baseline to differentiate against.

\subsection{Benchmark Saturation}

Our test environment is a flat network of hosts with common Microsoft Windows vulnerabilities, similar to a badly maintained SME network. In our runs, a minimal scaffold paired with a frontier model fully compromised all three GOAD domains in 3/5 runs for both tested models (Table~\ref{tab:performance}). Targets of this difficulty thus leave little headroom for measuring further progress. Harder testbeds could use hardened networks with fewer vulnerabilities, multi-stage network architectures, deploy additional active defenders, or provide a combination thereof. We designed Cochise with an easily changeable scenario prompt to allow quick adaptation to such new target environments.

\subsection{Ethics}

Any offensive security research is inherently dual-use research of concern (DURC). In our own line of research, we found the improvement in the cybersecurity capabilities in of off-the-shelf LLMs between our original research work~\cite{10.1145/3766895} and this publication especially concerning. We release Cochise to improve the reproducibility and comparability of research on autonomous penetration-testing agents, not to provide an operational red-team assistant. Because the harness is minimal, the offensive capability it exposes stems from the underlying LLMs rather than from our prototype. To prevent abuse, their providers are increasingly choosing \textit{structured access} as a protective measure. The readability of our prototype also helps malicious adversaries, but these parties typically use off-the-shelf offensive tooling instead of building bespoke tools, so our work should primarily benefit researchers and defenders.

\section{Conclusion}

We present Cochise, a 630 LOC reference harness for autonomous penetration-testing experiments, together with replay and analysis tooling and a corpus of trajectory logs from a live Active Directory testbed. We discuss how a bespoke security harness structures trajectories for later analysis, and how it contributes to the safety of experiments. The source code and example trajectories are provided to aid development of custom agents and as a baseline to benchmark against.

\section{Data Availability Statement}

All tools, together with a data-set of example penetration-testing trajectories, are available in a public github repository at \url{https://github.com/andreashappe/cochise} under a permissive open-source license. All released cochise versions are automatically archived on Zenodo~\cite{happe_2026_21770094}.

\bibliographystyle{ACM-Reference-Format}
\bibliography{bibliography}

@article{10.1145/3766895,
author = {Happe, Andreas and Cito, J\"{u}rgen},
title = {Can LLMs Hack Enterprise Networks? Autonomous Assumed Breach Penetration-Testing Active Directory Networks},
year = {2025},
publisher = {Association for Computing Machinery},
address = {New York, NY, USA},
issn = {1049-331X},
url = {https://doi.org/10.1145/3766895},
doi = {10.1145/3766895},
note = {Just Accepted},
journal = {ACM Trans. Softw. Eng. Methodol.},
month = sep
}

@inproceedings{10.1145/3611643.3613083,
author = {Happe, Andreas and Cito, J\"{u}rgen},
title = {Getting pwn’d by AI: Penetration Testing with Large Language Models},
year = {2023},
isbn = {9798400703270},
publisher = {Association for Computing Machinery},
address = {New York, NY, USA},
url = {https://doi.org/10.1145/3611643.3613083},
doi = {10.1145/3611643.3613083},
booktitle = {Proceedings of the 31st ACM Joint European Software Engineering Conference and Symposium on the Foundations of Software Engineering},
pages = {2082–2086},
numpages = {5},
location = {San Francisco, CA, USA},
series = {ESEC/FSE 2023}
}

@article{happe2026llms,
  author       = {Andreas Happe and
                  Aaron Kaplan and
                  J{\"{u}}rgen Cito},
  title        = {LLMs as Hackers: Autonomous Linux Privilege Escalation Attacks},
  journal      = {Empir. Softw. Eng.},
  volume       = {31},
  number       = {3},
  pages        = {70},
  year         = {2026},
  url          = {https://doi.org/10.1007/s10664-025-10758-3},
  doi          = {10.1007/S10664-025-10758-3},
  bibsource    = {dblp computer science bibliography, https://dblp.org}
}

@misc{deng2026makesgoodllmagent,
      title={What Makes a Good LLM Agent for Real-world Penetration Testing?}, 
      author={Gelei Deng and Yi Liu and Yuekang Li and Ruozhao Yang and Xiaofei Xie and Jie Zhang and Han Qiu and Tianwei Zhang},
      year={2026},
      archivePrefix={arXiv},
      primaryClass={cs.CR},
      url={https://arxiv.org/abs/2602.17622}, 
}

@misc{mayoralvilches2025caiopenbugbountyready,
      title={CAI: An Open, Bug Bounty-Ready Cybersecurity AI}, 
      author={Víctor Mayoral-Vilches and Luis Javier Navarrete-Lozano and María Sanz-Gómez and Lidia Salas Espejo and Martiño Crespo-Álvarez and Francisco Oca-Gonzalez and Francesco Balassone and Alfonso Glera-Picón and Unai Ayucar-Carbajo and Jon Ander Ruiz-Alcalde and Stefan Rass and Martin Pinzger and Endika Gil-Uriarte},
      year={2025},
      archivePrefix={arXiv},
      primaryClass={cs.CR},
      url={https://arxiv.org/abs/2504.06017}, 
}

@misc{singer2025incalmoautonomousllmassistedred,
      title={Incalmo: An Autonomous LLM-assisted System for Red Teaming Multi-Host Networks}, 
      author={Brian Singer and Keane Lucas and Lakshmi Adiga and Meghna Jain and Lujo Bauer and Vyas Sekar},
      year={2025},
      eprint={2501.16466},
      archivePrefix={arXiv},
      primaryClass={cs.CR},
      url={https://arxiv.org/abs/2501.16466}, 
}

@inproceedings{yang2024sweagent,
  title={{SWE}-agent: Agent-Computer Interfaces Enable Automated Software Engineering},
  author={John Yang and Carlos E Jimenez and Alexander Wettig and Kilian Lieret and Shunyu Yao and Karthik R Narasimhan and Ofir Press},
  booktitle={The Thirty-eighth Annual Conference on Neural Information Processing Systems},
  year={2024},
  url={https://arxiv.org/abs/2405.15793}
}

@misc{yao2023reactsynergizingreasoningacting,
      title={ReAct: Synergizing Reasoning and Acting in Language Models}, 
      author={Shunyu Yao and Jeffrey Zhao and Dian Yu and Nan Du and Izhak Shafran and Karthik Narasimhan and Yuan Cao},
      year={2023},
      eprint={2210.03629},
      archivePrefix={arXiv},
      primaryClass={cs.CL},
      url={https://arxiv.org/abs/2210.03629}, 
}

@inproceedings{shinn2023reflexion,
  author       = {Noah Shinn and
                  Federico Cassano and
                  Ashwin Gopinath and
                  Karthik Narasimhan and
                  Shunyu Yao},
  editor       = {Alice Oh and
                  Tristan Naumann and
                  Amir Globerson and
                  Kate Saenko and
                  Moritz Hardt and
                  Sergey Levine},
  title        = {Reflexion: language agents with verbal reinforcement learning},
  booktitle    = {Advances in Neural Information Processing Systems 36: Annual Conference
                  on Neural Information Processing Systems 2023, NeurIPS 2023, New Orleans,
                  LA, USA, December 10 - 16, 2023},
  year         = {2023},
  url          = {http://papers.nips.cc/paper\_files/paper/2023/hash/1b44b878bb782e6954cd888628510e90-Abstract-Conference.html},
  bibsource    = {dblp computer science bibliography, https://dblp.org}
}

@misc{yang2026betterharnessessmallermodels,
      title={Better Harnesses, Smaller Models: Building 90\% Cheaper Agents via Automated Harness Adaptation}, 
      author={Chenyang Yang and Xinran Zhao and Tongshuang Wu and Christian Kästner},
      year={2026},
      eprint={2607.08938},
      archivePrefix={arXiv},
      primaryClass={cs.SE},
      url={https://arxiv.org/abs/2607.08938}, 
}

@misc{anthropic2026cybertoolkits,
  title = {AI models are showing a greater ability to find and exploit vulnerabilities on realistic cyber ranges},
  author = {{Anthropic}},
  year = {2026},
  month = {January},
  day = {16},
  url = {https://www.anthropic.com/research/cyber-toolkits-update},
  howpublished = {\url{https://www.anthropic.com/research/cyber-toolkits-update}},
  note = {Accessed: 2026-08-03}
}

@software{happe_2026_21770094,
  author       = {Happe, Andreas and
                  Cito, Jürgen},
  title        = {cochise},
  month        = aug,
  year         = 2026,
  publisher    = {Zenodo},
  version      = {v0.4.1},
  doi          = {10.5281/zenodo.21770094},
  url          = {https://doi.org/10.5281/zenodo.21770094},
}
\end{document}